
%
%
%
%
%
%
%
\magnification = 1200

\def\Resetstrings{
    \def\present{ }\let\bgroup={\let\egroup=}
    \def\Astr{}\def\astr{}\def\Atest{}\def\atest{}%
    \def\Bstr{}\def\bstr{}\def\Btest{}\def\btest{}%
    \def\Cstr{}\def\cstr{}\def\Ctest{}\def\ctest{}%
    \def\Dstr{}\def\dstr{}\def\Dtest{}\def\dtest{}%
    \def\Estr{}\def\estr{}\def\Etest{}\def\etest{}%
    \def\Fstr{}\def\fstr{}\def\Ftest{}\def\ftest{}%
    \def\Gstr{}\def\gstr{}\def\Gtest{}\def\gtest{}%
    \def\Hstr{}\def\hstr{}\def\Htest{}\def\htest{}%
    \def\Istr{}\def\istr{}\def\Itest{}\def\itest{}%
    \def\Jstr{}\def\jstr{}\def\Jtest{}\def\jtest{}%
    \def\Kstr{}\def\kstr{}\def\Ktest{}\def\ktest{}%
    \def\Lstr{}\def\lstr{}\def\Ltest{}\def\ltest{}%
    \def\Mstr{}\def\mstr{}\def\Mtest{}\def\mtest{}%
    \def\Nstr{}\def\nstr{}\def\Ntest{}\def\ntest{}%
    \def\Ostr{}\def\ostr{}\def\Otest{}\def\otest{}%
    \def\Pstr{}\def\pstr{}\def\Ptest{}\def\ptest{}%
    \def\Qstr{}\def\qstr{}\def\Qtest{}\def\qtest{}%
    \def\Rstr{}\def\rstr{}\def\Rtest{}\def\rtest{}%
    \def\Sstr{}\def\sstr{}\def\Stest{}\def\stest{}%
    \def\Tstr{}\def\tstr{}\def\Ttest{}\def\ttest{}%
    \def\Ustr{}\def\ustr{}\def\Utest{}\def\utest{}%
    \def\Vstr{}\def\vstr{}\def\Vtest{}\def\vtest{}%
    \def\Wstr{}\def\wstr{}\def\Wtest{}\def\wtest{}%
    \def\Xstr{}\def\xstr{}\def\Xtest{}\def\xtest{}%
    \def\Ystr{}\def\ystr{}\def\Ytest{}\def\ytest{}%
}
\Resetstrings

\def\Refformat{
         \if\Jtest\present
             {\if\Vtest\present\journalarticleformat
                  \else\conferencereportformat\fi}
            \else\if\Btest\present\bookarticleformat
               \else\if\Rtest\present\technicalreportformat
                  \else\if\Itest\present\bookformat
                     \else\otherformat\fi\fi\fi\fi}

\def\Rpunct{
   \def\Lspace{ }%
   \def\Lperiod{ }
   \def\Lcomma{ }
   \def\Lquest{ }
   \def\Lcolon{ }
   \def\Lscolon{ }
   \def\Lbang{ }
   \def\Lquote{ }
   \def\Lqquote{ }
   \def\Lrquote{ }
   \def\Rspace{}%
   \def\Rperiod{.}
   \def\Rcomma{,}
   \def\Rquest{?}
   \def\Rcolon{:}
   \def\Rscolon{;}
   \def\Rbang{!}
   \def\Rquote{'}
   \def\Rqquote{"}
   \def\Rrquote{`}
   }

\def\Lpunct{
   \def\Lspace{}%
   \def\Lperiod{\unskip.}
   \def\Lcomma{\unskip,}
   \def\Lquest{\unskip?}
   \def\Lcolon{\unskip:}
   \def\Lscolon{\unskip;}
   \def\Lbang{\unskip!}
   \def\Lquote{\unskip'}
   \def\Lqquote{\unskip"}
   \def\Lrquote{\unskip`}
   \def\Rspace{\spacefactor=1000}%
   \def\Rperiod{\spacefactor=3000}
   \def\Rcomma{\spacefactor=1250}
   \def\Rquest{\spacefactor=3000}
   \def\Rcolon{\spacefactor=2000}
   \def\Rscolon{\spacefactor=1250}
   \def\Rbang{\spacefactor=3000}
   \def\Rquote{\spacefactor=1000}
   \def\Rqquote{\spacefactor=1000}
   \def\Rrquote{\spacefactor=1000}
   }

\def\Smallcapsaand{
     \def\Aand{\unskip\bgroup{\Smallcapsfont\ AND }\egroup}%
     \def\Aandd{\unskip\bgroup{\Smallcapsfont\ AND }\egroup}%
     \def\eand{\unskip\bgroup\Smallcapsfont\ AND \egroup}%
     \def\eandd{\unskip\bgroup\Smallcapsfont\ AND \egroup}%
   }

\def\Smallcapseand{
     \def\Eand{\unskip\bgroup\Smallcapsfont\ AND \egroup}%
     \def\Eandd{\unskip\bgroup\Smallcapsfont\ AND \egroup}%
     \def\aand{\unskip\bgroup\Smallcapsfont\ AND \egroup}%
     \def\aandd{\unskip\bgroup\Smallcapsfont\ AND \egroup}%
   }

\def\Refstda{
    \chardef\Ampersand=`\&
    \def\Acomma{\unskip, }
    \def\Aand{\unskip\ \Ampersand\ }
    \def\Aandd{\unskip\ \Ampersand\ }
    \def\Ecomma{\unskip, }
    \def\Eand{\unskip\ \Ampersand\ }
    \def\Eandd{\unskip\ \Ampersand\ }
    \def\acomma{\unskip, }
    \def\aand{\unskip\ \Ampersand\ }
    \def\aandd{\unskip\ \Ampersand\ }
    \def\ecomma{\unskip, }
    \def\eand{\unskip\ \Ampersand\ }
    \def\eandd{\unskip\ \Ampersand\ }
    \def\Namecomma{\unskip, }
    \def\Nameand{\unskip\ \Ampersand\ }
   \def\Nameandd{\unskip\ \Ampersand\ }
    \def\Revcomma{\unskip, }
    \def\Initper{.\ }
    \def\Initgap{\dimen0=\spaceskip\divide\dimen0 by 2\hskip-\dimen0}%
  }

   \def\Citefont{}
   \def\ACitefont{}
   \def\Authfont{}
   \def\Titlefont{}
   \def\Tomefont{\sl}
   \def\Volfont{}
   \def\Flagfont{}
   \def\Reffont{\rm}
   \def\Smallcapsfont{\sevenrm}
   \def\Flagstyle#1{\hangindent\parindent\indent\hbox to0pt
       {\hss[{\Flagfont#1}]\kern.5em}\ignorespaces}


\def\Citebrackets{\Rpunct
   \def\Lcitemark{\def\Cfont{\Citefont}[\bgroup\Cfont}
   \def\Rcitemark{\egroup]}
   \def\LAcitemark{\def\Cfont{\ACitefont}\bgroup\ACitefont}%
   \def\RAcitemark{\egroup}
   \def\LIcitemark{\egroup}
   \def\RIcitemark{\bgroup\Cfont}
   \def\Citehyphen{\egroup--\bgroup\Cfont}
   \def\Citecomma{\egroup,\hskip0pt\bgroup\Cfont}%
   \def\Citebreak{}
   }

\def\Citeparen{\Rpunct
   \def\Lcitemark{\def\Cfont{\Citefont}(\bgroup\Cfont}
   \def\Rcitemark{\egroup)}
   \def\LAcitemark{\def\Cfont{\ACitefont}\bgroup\ACitefont}%
   \def\RAcitemark{\egroup}
   \def\LIcitemark{\egroup}
   \def\RIcitemark{\bgroup\Cfont}
   \def\Citehyphen{\egroup--\bgroup\Cfont}
   \def\Citecomma{\egroup,\hskip0pt\bgroup\Cfont}%
   \def\Citebreak{}
   }

\def\Citesuper{\Lpunct
   \def\Lcitemark{\def\Cfont{\Citefont}\raise1ex\hbox\bgroup\bgroup\Cfont}%
   \def\Rcitemark{\egroup\egroup}
   \def\LAcitemark{\def\Cfont{\ACitefont}\bgroup\ACitefont}%
   \def\RAcitemark{\egroup}
   \def\LIcitemark{\egroup\egroup}
   \def\RIcitemark{\raise1ex\hbox\bgroup\bgroup\Cfont}%
   \def\Citehyphen{\egroup--\bgroup\Cfont}
   \def\Citecomma{\egroup,\hskip0pt\bgroup%
      \Cfont}
   \def\Citebreak{}
   }

\def\Citenamedate{\Rpunct
   \def\Lcitemark{
      \def\Citebreak{\egroup\ [\bgroup\Citefont}
      \def\Citecomma{\egroup]; 
         \bgroup\let\uchyph=1\Citefont}(\bgroup\let\uchyph=1\Citefont}%
   \def\Rcitemark{\egroup])}
   \def\LAcitemark{
      \def\Citebreak{\egroup\ [\bgroup\Citefont}\def\Citecomma{\egroup], %
         \bgroup\ACitefont }\bgroup\let\uchyph=1\ACitefont}%
   \def\RAcitemark{\egroup]}
  \def\Citehyphen{\egroup--\bgroup\Citefont}
   \def\LIcitemark{\egroup}
   \def\RIcitemark{\bgroup\Citefont}
   }

\def\Flagstyle#1{\hangindent\parindent\indent\hbox
to0pt{\hss[{\Flagfont#1}]\kern.5em}}

\def\journalarticleformat{\Reffont\let\uchyph=1\parindent=1.25pc\def\Comma{}%

\sfcode`\.=1000\sfcode`\?=1000\sfcode`\!=1000\sfcode`\:=1000\sfcode`%
\;=1000\sfcode`\,=1000
                \par\vfil\penalty-200\vfilneg
      \if\Ftest\present\Flagstyle\Fstr\fi%
       \if\Atest\present\bgroup\Authfont\Astr\egroup\def\Comma{\unskip, }\fi%
        \if\Ttest\present\Comma\bgroup``\Titlefont\Tstr\egroup\def\Comma{,"
}\fi%

\if\etest\present\if\Ttest\present{"}\fi%
\hskip.16667em(\bgroup\estr\egroup)\def\Comma{\unskip, }\fi%
          \if\Jtest\present\Comma\bgroup\Tomefont\Jstr\/\egroup\def\Comma{,
}\fi%

\if\Vtest\present\if\Jtest\present\hskip.16667em%
\else\Comma\fi\bgroup\Volfont\Vstr\egroup\def\Comma{, }\fi%
            \if\Dtest\present\hskip.16667em(\bgroup\Dstr\egroup)\def\Comma{,
}\fi%
             \if\Ptest\present\bgroup, \Pstr\egroup\def\Comma{, }\fi%

\if\ttest\present\Comma\bgroup``\Titlefont\tstr\egroup\def\Comma{," }\fi%

\if\jtest\present\Comma\bgroup\Tomefont\jstr\/\egroup\def\Comma{, }\fi%

\if\vtest\present\if\jtest\present\hskip.16667em%
\else\Comma\fi\bgroup\Volfont\vstr\egroup\def\Comma{, }\fi%

\if\dtest\present\hskip.16667em(\bgroup\dstr\egroup)\def\Comma{, }\fi%
                  \if\ptest\present\bgroup, \pstr\egroup\def\Comma{, }\fi%
                   \if\Gtest\present{\Comma Gov't ordering no.
}\bgroup\Gstr\egroup\def\Comma{, }\fi%
                    \if\Otest\present{\Comma\bgroup\Ostr\egroup.}\else{.}\fi%
                     \vskip3ptplus1ptminus1pt}

\def\conferencereportformat{\Reffont\let\uchyph=1\parindent=1.25pc\def\Comma{}%

\sfcode`\.=1000\sfcode`\?=1000\sfcode`\!=1000\sfcode`\:=1000\sfcode`%
\;=1000\sfcode`\,=1000
                \par\vfil\penalty-200\vfilneg
      \if\Ftest\present\Flagstyle\Fstr\fi%
       \if\Atest\present\bgroup\Authfont\Astr\egroup\def\Comma{\unskip, }\fi%
        \if\Ttest\present\Comma\bgroup``\Titlefont\Tstr\egroup\def\Comma{,"
}\fi%
         \if\Jtest\present\Comma\bgroup\Tomefont\Jstr\/\egroup\def\Comma{,
}\fi%
          \if\Ctest\present\Comma\bgroup\Cstr\egroup\def\Comma{, }\fi%
           \if\Dtest\present\hskip.16667em(\bgroup\Dstr\egroup)\def\Comma{,
}\fi%
            \if\Otest\present{\Comma\bgroup\Ostr\egroup.}\else{.}\fi%
             \vskip3ptplus1ptminus1pt}

\def\bookarticleformat{\Reffont\let\uchyph=1\parindent=1.25pc\def\Comma{}%

\sfcode`\.=1000\sfcode`\?=1000\sfcode`\!=1000\sfcode`\:=1000\sfcode`%
\;=1000\sfcode`\,=1000
                \par\vfil\penalty-200\vfilneg
      \if\Ftest\present\Flagstyle\Fstr\fi%
       \if\Atest\present\bgroup\Authfont\Astr\egroup\def\Comma{\unskip, }\fi%
        \if\Ttest\present\Comma\bgroup``\Titlefont\Tstr\egroup\def\Comma{,"
}\fi%

\if\etest\present\if\Ttest\present"\fi%
\hskip.16667em(\bgroup\estr\egroup)\def\Comma{\unskip, }\fi%
          \if\Btest\present\Comma in
\bgroup\Tomefont\Bstr\/\egroup\def\Comma{\unskip, }\fi%
           \if\otest\present\ \bgroup\ostr\egroup\def\Comma{, }\fi%
            \if\Etest\present\Comma\bgroup\Estr\egroup\unskip,
\ifnum\Ecnt>1eds.\else ed.\fi\def\Comma{, }\fi%
             \if\Stest\present\Comma\bgroup\Sstr\egroup\def\Comma{, }\fi%

\if\Vtest\present\bgroup\hskip.16667em\#\Volfont\Vstr\egroup\def\Comma{, }\fi%

\if\Ntest\present\bgroup\hskip.16667em\#\Volfont\Nstr\egroup\def\Comma{, }\fi%
                \if\Itest\present\Comma\bgroup\Istr\egroup\def\Comma{, }\fi%
                 \if\Ctest\present\Comma\bgroup\Cstr\egroup\def\Comma{, }\fi%
                  \if\Dtest\present\Comma\bgroup\Dstr\egroup\def\Comma{, }\fi%
                   \if\Ptest\present\Comma\Pstr\def\Comma{, }\fi%

\if\ttest\present\Comma\bgroup``\Titlefont\Tstr\egroup\def\Comma{," }\fi%
                     \if\btest\present\Comma in
\bgroup\Tomefont\bstr\egroup\def\Comma{, }\fi%
                       \if\atest\present\Comma\bgroup\astr\egroup\unskip,
\if\acnt\present eds.\else ed.\fi\def\Comma{, }\fi%
                        \if\stest\present\Comma\bgroup\sstr\egroup\def\Comma{,
}\fi%

\if\vtest\present\bgroup\hskip.16667em\#\Volfont\vstr\egroup\def\Comma{, }\fi%

\if\ntest\present\bgroup\hskip.16667em\#\Volfont\nstr\egroup\def\Comma{, }\fi%

\if\itest\present\Comma\bgroup\istr\egroup\def\Comma{, }\fi%

\if\ctest\present\Comma\bgroup\cstr\egroup\def\Comma{, }\fi%

\if\dtest\present\Comma\bgroup\dstr\egroup\def\Comma{, }\fi%
                              \if\ptest\present\Comma\pstr\def\Comma{, }\fi%
                               \if\Gtest\present{\Comma Gov't ordering no.
}\bgroup\Gstr\egroup\def\Comma{, }\fi%

\if\Otest\present{\Comma\bgroup\Ostr\egroup.}\else{.}\fi%
                                 \vskip3ptplus1ptminus1pt}

\def\bookformat{\Reffont\let\uchyph=1\parindent=1.25pc\def\Comma{}%

\sfcode`\.=1000\sfcode`\?=1000\sfcode`\!=1000\sfcode`\:=1000%
\sfcode`\;=1000\sfcode`\,=1000
                \par\vfil\penalty-200\vfilneg
      \if\Ftest\present\Flagstyle\Fstr\fi%
       \if\Atest\present\bgroup\Authfont\Astr\egroup\def\Comma{\unskip, }%
\else\if\Etest\present\bgroup\def\Eand{\Aand}\def\Eandd{\Aandd}%
\Authfont\Estr\egroup\unskip, \ifnum\Ecnt>1eds.\else ed.\fi\def\Comma{, }%

\else\if\Itest\present\bgroup\Authfont\Istr\egroup\def\Comma{, }\fi\fi\fi%

\if\Ttest\present\Comma\bgroup\Tomefont\Tstr\/\egroup\def\Comma{\unskip, }%

\else\if\Btest\present\Comma\bgroup\Tomefont\Bstr\/\egroup\def\Comma{\unskip,
}\fi\fi%
            \if\otest\present\ \bgroup\ostr\egroup\def\Comma{, }\fi%

\if\etest\present\hskip.16667em(\bgroup\estr\egroup)\def\Comma{\unskip, }\fi%
              \if\Stest\present\Comma\bgroup\Sstr\egroup\def\Comma{, }\fi%

\if\Vtest\present\bgroup\hskip.16667em\#\Volfont\Vstr\egroup\def\Comma{, }\fi%

\if\Ntest\present\bgroup\hskip.16667em\#\Volfont\Nstr\egroup\def\Comma{, }\fi%
                 \if\Atest\present\if\Itest\present
                         \Comma\bgroup\Istr\egroup\def\Comma{\unskip, }\fi%
                      \else\if\Etest\present\if\Itest\present
                              \Comma\bgroup\Istr\egroup\def\Comma{\unskip,
}\fi\fi\fi%
                     \if\Ctest\present\Comma\bgroup\Cstr\egroup\def\Comma{,
}\fi%
                      \if\Dtest\present\Comma\bgroup\Dstr\egroup\def\Comma{,
}\fi%

\if\ttest\present\Comma\bgroup\Tomefont\tstr\egroup\def\Comma{, }%

\else\if\btest\present\Comma\bgroup\Tomefont\bstr\egroup\def\Comma{, }\fi\fi%

\if\stest\present\Comma\bgroup\sstr\egroup\def\Comma{, }\fi%

\if\vtest\present\bgroup\hskip.16667em\#\Volfont\vstr\egroup\def\Comma{, }\fi%

\if\ntest\present\bgroup\hskip.16667em\#\Volfont\nstr\egroup\def\Comma{, }\fi%

\if\itest\present\Comma\bgroup\istr\egroup\def\Comma{, }\fi%

\if\ctest\present\Comma\bgroup\cstr\egroup\def\Comma{, }\fi%

\if\dtest\present\Comma\bgroup\dstr\egroup\def\Comma{, }\fi%
                                \if\Gtest\present{\Comma Gov't ordering no.
}\bgroup\Gstr\egroup\def\Comma{, }\fi%

\if\Otest\present{\Comma\bgroup\Ostr\egroup.}\else{.}\fi%
                                  \vskip3ptplus1ptminus1pt}

\def\technicalreportformat{\Reffont\let\uchyph=1\parindent=1.25pc\def\Comma{}%
\sfcode`\.=1000\sfcode`\?=1000\sfcode`\!=1000\sfcode`\:=1000\sfcode`%
\;=1000\sfcode`\,=1000
                \par\vfil\penalty-200\vfilneg
      \if\Ftest\present\Flagstyle\Fstr\fi%
       \if\Atest\present\bgroup\Authfont\Astr\egroup\def\Comma{\unskip, }%
\else\if\Etest\present\bgroup\def\Eand{\Aand}%
\def\Eandd{\Aandd}\Authfont\Estr\egroup\unskip, \ifnum\Ecnt>1eds.%
\else ed.\fi\def\Comma{, }%

\else\if\Itest\present\bgroup\Authfont\Istr\egroup\def\Comma{, }\fi\fi\fi%
          \if\Ttest\present\Comma\bgroup``\Titlefont\Tstr\egroup\def\Comma{,"
}\fi%
           \if\Atest\present\if\Itest\present
                   \Comma\bgroup\Istr\egroup\def\Comma{, }\fi%
                \else\if\Etest\present\if\Itest\present
                        \Comma\bgroup\Istr\egroup\def\Comma{, }\fi\fi\fi%
            \if\Rtest\present\Comma\bgroup\Rstr\egroup\def\Comma{, }\fi%
             \if\Ctest\present\Comma\bgroup\Cstr\egroup\def\Comma{, }\fi%
              \if\Dtest\present\Comma\bgroup\Dstr\egroup\def\Comma{, }\fi%

\if\ttest\present\Comma\bgroup``\Titlefont\tstr\egroup\def\Comma{," }\fi%
                \if\itest\present\Comma\bgroup\istr\egroup\def\Comma{, }\fi%
                 \if\rtest\present\Comma\bgroup\rstr\egroup\def\Comma{, }\fi%
                  \if\ctest\present\Comma\bgroup\cstr\egroup\def\Comma{, }\fi%
                   \if\dtest\present\Comma\bgroup\dstr\egroup\def\Comma{, }\fi%
                    \if\Gtest\present{\Comma Gov't ordering no.
}\bgroup\Gstr\egroup\def\Comma{, }\fi%
                     \if\Otest\present{\Comma\bgroup\Ostr\egroup.}\else{.}\fi%
                      \vskip3ptplus1ptminus1pt}

\def\otherformat{\Reffont\let\uchyph=1\parindent=1.25pc\def\Comma{}%
\sfcode`\.=1000\sfcode`\?=1000\sfcode`\!=1000\sfcode`\:=1000%
\sfcode`\;=1000\sfcode`\,=1000
                \par\vfil\penalty-200\vfilneg
      \if\Ftest\present\Flagstyle\Fstr\fi%
       \if\Atest\present\bgroup\Authfont\Astr\egroup\def\Comma{\unskip, }%
\else\if\Etest\present\bgroup\def\Eand{\Aand}\def\Eandd{\Aandd}%
\Authfont\Estr\egroup\unskip, \ifnum\Ecnt>1eds.\else ed.\fi\def\Comma{, }%

\else\if\Itest\present\bgroup\Authfont\Istr\egroup\def\Comma{, }\fi\fi\fi%
          \if\Ttest\present\Comma\bgroup``\Titlefont\Tstr\egroup\def\Comma{,"
}\fi%
            \if\Atest\present\if\Itest\present
                    \Comma\bgroup\Istr\egroup\def\Comma{, }\fi%
                 \else\if\Etest\present\if\Itest\present
                         \Comma\bgroup\Istr\egroup\def\Comma{, }\fi\fi\fi%
                 \if\Ctest\present\Comma\bgroup\Cstr\egroup\def\Comma{, }\fi%
                  \if\Dtest\present\Comma\bgroup\Dstr\egroup\def\Comma{, }\fi%
                   \if\Gtest\present{\Comma Gov't ordering no.
}\bgroup\Gstr\egroup\def\Comma{, }\fi%
                    \if\Otest\present{\Comma\bgroup\Ostr\egroup.}\else{.}\fi%
                     \vskip3ptplus1ptminus1pt}
\Refstda\Citebrackets 

\documentstyle{amsppt}
\topmatter
\title A three parameter invariant of oriented links\endtitle
\author Bruce W. Westbury \endauthor
\date 22 April 1994 \enddate
\address
Department of Mathematics
University of Nottingham
University Park
Nottingham
NG7 2RD
\endaddress
\abstract This paper defines a new sequence of finite dimensional
algebras as quotients of the group algebras of the braid groups.
This sequence depends on three homogeneous parameters and has a
one-parameter family of Markov traces, and so gives a three parameter
invariant of oriented links.
\endabstract
\email bww\@uk.ac.nott.maths \endemail
\endtopmatter

\document
\head Introduction \endhead
The Jones polynomial was originaly constructed by defining a Markov
trace on the Temperley-Lieb algebras, see\Lspace \Lcitemark 1\Rcitemark
\Rspace{}.
Similarly the HOMFLY polynomial can be constructed by defining a one parameter
family of Markov traces on the Hecke algebras, see\Lspace \Lcitemark
2\Rcitemark \Rspace{},
and the Kauffman polynomial can be constructed by defining a Markov trace on
the Birman-Wenzl-Murakami algebras, see\Lspace \Lcitemark 3\Rcitemark \Rspace{}
and\Lspace \Lcitemark 4\Rcitemark \Rspace{}.

In this paper we define a link invariant
by the same construction. First we define a tower of algebras,
$A_n$, for $n>0$. For each $n$, $A_n$ is a finite dimensional quotient
of the group algebra of the braid group on $n$ strings. Then we show
that there is a one-parameter family of Markov traces on this tower of
algebras.

The image of each of the standard braid group generators in
$A_n$ satisfies a fixed cubic relation and so has three eigenvalues.
These three eigenvalues are essentially arbitrary and are homogeneous
coordinates; so each $A_n$ is a two-dimensional family of algebras.
Since there is a one-parameter family of Markov traces this means that
the link invariant depends on three parameters. However the invariant
is a rational function and not a Laurent polynomial.

This link invariant has two specialisations to the Jones polynomial.
This link invariant also specialises to the link polynomial associated
to the spin 1, or adjoint, representation of $SU(2)$. Since this
representation is also the fundamental representation of $SO(3)$
this specialisation is also a specialisation of the Kauffman polynomial.

\head A tower of algebras \endhead
In this section we define the sequence of algebras, $A_n$, for $n>0$.
For each $n>0$, $A_n$ is defined as a quotient of the group algebra
of the braid group on $n$ strings, $B_n$.

Each of these algebras is a finite dimensional unital algebra over the field
of homogeneous rational functions in the indeterminates $x$, $y$ and $z$
of degree zero which are invariant under the involution
$x\leftrightarrow x^{-1}$, $y\leftrightarrow y^{-1}$,
$z\leftrightarrow z^{-1}$.

The algebra $A_1$ is one dimensional. The algebra $A_2$ has dimension
three and is defined to be the quotient of the group algebra of $B_2$
by the relation
$$(\sigma_1-x)(\sigma_1-y)(\sigma_1-z)=0\tag1$$

The algebra $A_3$, as an abstract algebra, is the direct sum of five
matrix algebras of ranks $1$, $1$, $2$, $2$ and $3$. The dimension of
$A_3$ is therefore 19. The homomorphism $B_3\to A_3$ is defined by
specifying five irreducible representations of $B_3$ of these
dimensions.

The two one dimensional representations are $\sigma_i\mapsto x$ and
$\sigma_i\mapsto z$, for $i=1,2$.

The two two dimensional representations are
$$\alignedat 2
\sigma_1&\mapsto\pmatrix x&a_1\\0&y\endpmatrix &
\sigma_2&\mapsto\pmatrix y&0\\a_1^\prime&x\endpmatrix \\
\sigma_1&\mapsto\pmatrix z&a_2\\0&y\endpmatrix &
\sigma_2&\mapsto\pmatrix y&0\\a_2^\prime&z\endpmatrix
\endalignedat\tag2$$
where the constants are chosen to satisfy $a_1a_1^\prime=-xy$ and
$a_2a_2^\prime=-yz$, and the representations are, up to equivalence,
independent of this choice.

The three dimensional representation is
$$\sigma_1\mapsto\pmatrix x&b_1&{b_1b_2y\over y^2+xz}\\
0&y&b_2\\ 0&0&z\endpmatrix
\qquad
\sigma_2\mapsto\pmatrix z&0&0\\ b_1^\prime&y&0\\
{b_1^\prime b_2^\prime y\over y^2+xz}&b_2^\prime&z\endpmatrix\tag3$$
where the constants are chosen to satisfy
$b_1b_1^\prime=b_2b_2^\prime=-y^2-xz$ and the representation is,
up to equivalence, independent of this choice.

This completes the definition of $A_3$. Next we find generators and
defining relations.

The image of each of $\sigma_1$ and $\sigma_2$ in $A_3$ satisfies the
cubic relation $(1)$. Hence each of these
can be written as a linear combination of three orthogonal idempotents
which sum to the identity. Define $e_i$ and $g_i$, for $i=1,2$, by
$$
e_i={(\sigma_i-y)(\sigma_i-z)\over (x-y)(x-z)}
\qquad\text{and}\qquad
g_i={(\sigma_i-x)(\sigma_i-y)\over (z-x)(z-y)}
\tag4$$

Then, for $i=1,2$, $e_i$ and $g_i$ are orthogonal idempotents. Since
$$\aligned
\sigma_i&=y+(x-y)e_i+(z-y)g_i\\
\sigma_i^{-1}&=y^{-1}+(x^{-1}-y^{-1})e_i+(z^{-1}-y^{-1})g_i
\endaligned\tag5$$
the algebra $A_3$ is generated by $1,e_1,g_1,e_2,g_2$.

\comment
There are three symmetries of order two: an involution given by
$e_1\leftrightarrow e_2$ and $f_1\leftrightarrow f_2$; an involution
given by $x\leftarrow z$ and $e_i\leftrightarrow g_i$; and an
anti-involution given by $e_i\leftrightarrow e_i$ and
$g_i\leftrightarrow g_i$.
\endcomment

There are 13 words in these generators of length at most two, and 8
words of the form $abc$ where each of $a$ and $c$ is $e_1$ or $f_1$
and $b$ is $e_2$ or $f_2$. This gives a combined set of 21 words; since
the dimension of $A_3$ is 19 these must satisfy at least two linear
relations. These 21 words satisfy the two linear relations
$$\aligned
x(y^2-z^2)g_1e_2e_1&=z(x^2-y^2)g_1g_2e_1\\
e_1e_2g_1&=g_1e_2e_1
\endaligned\tag6$$
Note that it is sufficient to check these two relations in the
representation $(3)$ since both sides of both equations are zero in the
other four representations.

Now we show that there are no other linear relations among these 21
elements. The words $1$, $e_1$, $e_2$, $e_1e_2$ and $e_2e_1$ are linearly
independent in the two representations in which $g_i=0$, and similarly,
the words $1$, $g_1$, $g_2$, $g_1g_2$ and $g_2g_1$ are linearly
independent in the two representations in which $e_i=0$. Therefore, it
is sufficient to check that the remaining 12 words in the representation
$(3)$ span the vector space of $3\times 3$ matrices. This can be verified
by a direct calculation.

It follows from this that each of the 8 words of the form $abc$ where
each of $a$ and $c$ is $e_2$ or $f_2$ and $b$ is $e_1$ or $f_1$ can be
written as a linear combination of the above 21 words. These relations,
together with ones already given, are defining relations for the algebra
$A_3$. These relations can be determined explicitly, but are not
reproduced here as they are complicated and not particularly illuminating.
In fact, defining relations are the braid relations and two further
relations; one of these relations is given in\Lspace \Lcitemark 5\Rcitemark
\Rspace{} and the other
can be obtained from this by the involution $e_i\leftrightarrow g_i$
and $x\leftrightarrow y$. However the important feature of these relations
is that they show that
$$A_3=A_2+A_2e_2A_2+A_2g_2A_2\tag7$$

This ends the discussion for $n=3$. Next we discuss the general case.

For $n>3$, the algebra $A_n$ is generated by elements $1$,
$e_1,e_2,\ldots e_{n-1}$ and $g_1,g_2,\ldots ,g_{n-1}$.
For $1\leqslant i\leqslant n-1$, $e_i$ and $g_i$ are orthogonal idempotents;
for $|i-j|>1$, $e_ie_j=e_je_i$, $e_ig_j=g_je_i$ and $g_ig_j=g_jg_i$; and
for $1\leqslant i\leqslant n-2$ there is a monomorphism $A_3\to A_n$
with $e_1\mapsto e_i$, $g_1\mapsto g_i$, $e_2\mapsto e_{i+1}$ and
$g_2\mapsto g_{i+1}$.

The first property of these algebras is that for $n>1$,
$$A_{n+1}=A_n+A_ne_nA_n+A_ng_nA_n\tag8$$
This follows by a standard inductive argument on $n$. The result $(7)$
is the basis for the induction and is used in the inductive step.
In particular this shows that, for $n>1$, $A_n$ is finite dimensional;
more precisely, it follows that
$$\dim (A_n)\leqslant \prod_{k=1}^n(2^k-1)\tag9$$

\head Markov traces \endhead

In this section we find the general Markov trace on the tower of algebras
$A_n$. Let $\delta$ and $Z$ be new indeterminates. A Markov trace
consists of a trace map $\tau_n$ on $A_n$, for each $n>0$, such that,
for all $a\in A_n$,
$$\aligned
\tau_{n+1}(1)&=\delta^{n+1}\\
\tau_{n+1}(a)&=\delta\tau_n(a)\\
\tau_{n+1}(a\sigma_n^{\pm 1})&=Z^{\pm 1}\tau_n(a)
\endaligned\tag10$$

First we find $\tau_3$. The matrix trace of each irreducible representation
of $A_3$ is a trace on $A_3$, and any trace map can be written uniquely as
a linear combination of these five traces. Using $(7)$ and the relations
$(\sigma_1\sigma_2)e_1(\sigma_1\sigma_2)^-1=e_2$ and
$(\sigma_1\sigma_2)g_1(\sigma_1\sigma_2)^-1=g_2$, a trace is determined
by its values on the elements 1, $e_1$, $g_1$, $e_1g_2$, $e_1e_2$ and
$g_1g_2$.
Since there are six of these elements, these six values are not arbitrary
but must satisfy one linear relation.

The values of the trace, $\tau_3$, on these six elements are uniquely
determined by $(10)$ and are
$$\aligned
\tau_3(1)&=\delta^3\\
\tau_3(e_1)&=\delta {x(Z-(y+z)\delta+yzZ^{-1})\over (x-y)(x-z)}\\
\tau_3(g_1)&=\delta {z(Z-(x+y)\delta+xyZ^{-1})\over (z-y)(z-x)}\\
\tau_3(e_1g_2)&=\left({x(Z-(y+z)\delta+yzZ^{-1})\over (x-y)(x-z)}\right)
\left({z(Z-(x+y)\delta+xyZ^{-1})\over (z-y)(z-x)}\right)\\
\tau_3(e_1e_2)&=\left({x(Z-(y+z)\delta+yzZ^{-1})\over (x-y)(x-z)}\right)^2\\
\tau_3(g_1g_2)&=\left({z(Z-(x+y)\delta+xyZ^{-1})\over (z-y)(z-x)}\right)^2
\endaligned\tag11$$

Since there is one linear relation that these six values must satisfy
for $\tau_3$ to be a trace map we regard $Z$ as an additional parameter
and the linear relation as an equation for $\delta$. Any linear relation
will give a cubic equation for $\delta$, but a direct calculation gives
that the equation for $\delta$ is the following quadratic equation:
$$\gathered (x^3y+x^2yz-x^2z^2-xy^2z+xyz^2+yz^3)Z^2\delta^2\\
-(Z^2+xz)(2x^2y-x^2z-xy^2+xyz-xz^2-y^2z+2yz^2)Z\delta\\
+(x-y)(y-z)(x^2z^2+xzZ^2+Z^4)=0
\endgathered\tag12$$
Note that each of the coefficients is homogeneous of degree six and is also
invariant under $x\leftrightarrow z$, $y\leftrightarrow y$.

Define $\delta$ by $(12)$. It follows from $(8)$ that there is at most
one sequence of linear functionals, $\tau_n$, that satisfy
$$\aligned
\tau_{n+1}(a)&=\delta\tau_n(a)\\
\tau_{n+1}(ae_nb)&={x(Z-(y+z)\delta+yzZ^{-1})\over (x-y)(x-z)}\tau_n(ab)\\
\tau_{n+1}(ag_nb)&={z(Z-(x+y)\delta+xyZ^{-1})\over (z-y)(z-x)}\tau_n(ab)
\endaligned\tag13$$

In order to have a link invariant it remains to show that each of these
linear functionals is well-defined and that each of these linear functionals
is a trace map.

The reason it is not clear that these linear functionals are well-defined
is that the decomposition in $(8)$ is not a direct sum decomposition due
to the relations $(6)$.
In order to show these linear functionals are well-defined
it is sufficient to show that, for $a,b\in A_{n-1}$,
$$\aligned
x(y^2-z^2)\tau_{n+1}(ag_{n-1}e_ne_{n-1}b)
&=z(x^2-y^2)\tau_{n+1}(ag_{n-1}g_ne_b)\\
\tau_{n+1}(ae_{n-1}e_ng_b)&=\tau_{n+1}(ag_{n-1}e_ne_{n-1}b)
\endaligned\tag14$$
These equations are obviously satisfied since both sides of both equations
are zero by $(13)$.

The proof that each of these linear functionals is a trace follows by the
same argument as in\Lspace \Lcitemark 2\Rcitemark \Rspace{}. The proof is by
induction on $n$ and
uses the case $n=3$ as the basis of the induction and in the inductive
step.

To prove that $\tau_{n+1}$ is a trace it is sufficient to prove that
for all $a\in A_{n+1}$, $\tau_{n+1}(ae_n)=\tau_{n+1}(e_na)$ and
$\tau_{n+1}(ag_n)=\tau_{n+1}(g_na)$. Using $(8)$, there are three cases
to consider; $a\in A_n$, $a\in A_ne_nA_n$ and $A_ng_nA_n$. In each of
these cases use $(8)$ again to write each $A_n$ as the sum of $A_{n-1}$,
$A_{n-1}e_{n-1}A_{n-1}$ and $A_{n-1}g_{n-1}A_{n-1}$. This gives a large
number of special cases to consider, but each of these cases gives the
same equation as for the same special case with $n=3$. Since $\tau_3$
is a trace it follows that each of these equations is satisfied and so
$\tau_{n+1}$ is a trace.

If $\beta$ is a braid on $n$ strings with writhe $w(\beta)$ and with oriented
link closure $\hat\beta$, then define $L(\hat\beta)$ by
$$L(\hat\beta)=Z^{-w(\beta)}\tau_n(\beta)\tag15$$
Then by Markov's theorem, $L(\hat\beta)$ is an invariant of the isotopy
class of the oriented link $\hat\beta$.

The invariant $L(\hat\beta)$ is not a Laurent polynomial but is a rational
function. Since $\delta$ satisfies a quadratic relation, $L(\hat\beta)$
can be written uniquely as $L_0(\hat\beta)+\delta L_1(\hat\beta)$. Each
of the functions $L_0(\hat\beta)$ and $L_1(\hat\beta)$ is a homogeneous
rational function in $x$, $y$, $z$ and $Z$ of degree 0, and is a Laurent
polynomial in $Z$ with coefficients which are homogeneous rational
functions in $x$, $y$ and $z$.

There are two epimorphisms from the tower of algebras $A_n$ to the tower
of Temperley-Lieb algebras one given by $e_i=0$ and the other by $g_i=0$.
Hence there are two different specialisations of $L$ to a Jones
polynomial.

This link invariant also specialises to the link polynomial associated
to the spin 1, or adjoint, representation of $SU(2)$. The composition
factors of the tensor product of three copies of the spin 1
representation are the representations with spin 0,1 2 and 3 with
multiplicities 1,3,2 and 1 respectively. This gives representations
of the braid group on three strings of these dimensions and it is
sufficient to check that they are irreducible and isomorphic to ones
defined in $(2)$ and $(3)$.

Finally, $L$ specialises to the link invariant defined in\Lspace \Lcitemark
5\Rcitemark \Rspace{}.
In this specialisation, $x$, $y$ and $z$ satisfy $y^2=\omega xz$ where
$\omega$ is a primitive cube root of 1, and this link invariant has the
curious feature that $D=0$.

These properties show the similarities between $L$ and other known link
invariants. There are other two variables link polynomials known with
some of these properties, for example, the HOMFLY polynomial, the
Kauffman polynomial and the two-cabling of the HOMFLY polynomial. The
link invariant $L$ does not specialise to any of these three polynomials.
An elementary, but complicated, way to show this is to show that the
relation $(12)$ is not satisfied, in each case.

\head References \endhead

\message{REFERENCE LIST}

\bgroup\Resetstrings%
\def\Ecnt{0}\def\acnt{0}%
\def\Ftest{ }\def\Fstr{1}%
\def\Atest{ }\def\Astr{Vaughan F. R. Jones}%
\def\Ttest{ }\def\Tstr{Index for subfactors}%
\def\Jtest{ }\def\Jstr{Invent. Math.}%
\def\Dtest{ }\def\Dstr{1983}%
\def\Vtest{ }\def\Vstr{72}%
\def\Ptest{ }\def\Pstr{1--25}%
\Refformat\egroup%

\bgroup\Resetstrings%
\def\Ecnt{0}\def\acnt{0}%
\def\Ftest{ }\def\Fstr{2}%
\def\Atest{ }\def\Astr{Vaughan F. R. Jones}%
\def\Ttest{ }\def\Tstr{Hecke algebra representations of braid groups and link
polynomials}%
\def\Jtest{ }\def\Jstr{Ann. of Math. (2)}%
\def\Vtest{ }\def\Vstr{126}%
\def\Dtest{ }\def\Dstr{1987}%
\def\Ptest{ }\def\Pstr{335--388}%
\Refformat\egroup%

\bgroup\Resetstrings%
\def\Ecnt{0}\def\acnt{0}%
\def\Ftest{ }\def\Fstr{3}%
\def\Atest{ }\def\Astr{Joan S. Birman%
  \Aand Hans Wenzl}%
\def\Ttest{ }\def\Tstr{Braids, link polynomials and a new algebra}%
\def\Jtest{ }\def\Jstr{Trans. Amer. Math. Soc.}%
\def\Dtest{ }\def\Dstr{1989}%
\def\Ptest{ }\def\Pstr{249--273}%
\def\Vtest{ }\def\Vstr{313}%
\Refformat\egroup%

\bgroup\Resetstrings%
\def\Ecnt{0}\def\acnt{0}%
\def\Ftest{ }\def\Fstr{4}%
\def\Atest{ }\def\Astr{J. Murakami}%
\def\Ttest{ }\def\Tstr{The Kauffman polynomial of links and invariant theory}%
\def\Jtest{ }\def\Jstr{Osaka J. Math.}%
\def\Vtest{ }\def\Vstr{24}%
\def\Dtest{ }\def\Dstr{1987}%
\def\Ptest{ }\def\Pstr{745--758}%
\Refformat\egroup%

\bgroup\Resetstrings%
\def\Ecnt{0}\def\acnt{0}%
\def\Ftest{ }\def\Fstr{5}%
\def\Atest{ }\def\Astr{Mo-Lin Ge%
  \Acomma Guang-Chun Liu%
  \Acomma Chang-Pu Sun%
  \Aandd Yi-Wen Wang}%
\def\Ttest{ }\def\Tstr{Non-Birman-Wenzl algebraic properties and redundancy of
exotic enhanced Yang-Baxter operator for spin model}%
\def\Dtest{ }\def\Dstr{1994}%
\def\Ptest{ }\def\Pstr{393--404}%
\def\Vtest{ }\def\Vstr{27}%
\def\Jtest{ }\def\Jstr{J. Phys. A}%
\Refformat\egroup%

\enddocument